\documentclass[%
 reprint, 
 amsmath,amssymb,
 aps,
pra,
floatfix,
]{revtex4-2}

\usepackage{graphicx}
\usepackage{epsfig} 
\usepackage{dcolumn}
\usepackage{bm}
\usepackage{epstopdf}
\usepackage{multirow}
\usepackage{amsmath}
\usepackage{booktabs}
\usepackage{color}
\usepackage{gensymb}
\usepackage{kantlipsum}
\usepackage{hyperref}
\usepackage{braket}
\usepackage{physics}
\usepackage{url}
\usepackage[utf8]{inputenc}
\usepackage[T1]{fontenc}
\usepackage{mathptmx}
\usepackage{lineno}
\usepackage[dvipsnames]{xcolor}
\usepackage[normalem]{ulem}
\usepackage{makecell}
\usepackage{bm}

\hypersetup{hidelinks}

\begin{document}

\title{A Mid-Infrared Platform Based on Strontium Tweezer Arrays}

\author{Aaron Holman$^{1}$}
\thanks{These authors contributed equally.}
\author{Ximo Sun$^{1}$}
\thanks{These authors contributed equally.}
\author{Bojeong Seo$^{1}$}
\author{Joshua Corn$^{1}$}
\author{Zezheng Zhu$^{2}$}
\author{Yuan Xu$^{2}$}
\author{Jiahao Wu$^{2}$}
\author{Nanfang Yu$^{2}$}
\author{Dmytro Filin$^{3}$}
\author{Marianna Safronova$^{3}$}
\author{Sebastian Will$^{1}$}
\email{sebastian.will@columbia.edu}
\affiliation{$^{1}$Department of Physics, Columbia University, New York, New York 10027, USA}
\affiliation{$^{2}$Department of Applied Physics and Applied Mathematics, Columbia University, New York, New York 10027, USA}
\affiliation{$^{3}$Department of Physics and Astronomy, University of Delaware, Newark, Delaware 19716, USA}

\date{\today}

\begin{abstract}
Subwavelength atomic tweezer arrays, in which atoms can be positioned at distances smaller than their emission wavelength, have been proposed as a versatile platform to study collective emission phenomena, such as superradiance and subradiance. Experimentally, the realization of such arrays has been a challenge as typical emission wavelengths in the visible or near-infrared are short compared to typical tweezer spacings in the micrometer range. Here, we use $^{88}$Sr atoms in optical tweezer arrays to access a mid-infrared transition at 2,923 nm ($5s5p\:^{3}P_{2} \rightarrow\, 5s4d\:^{3}D_{3}$). We identify a magic trapping wavelength at 597.14(3) nm and demonstrate single-atom preparation and imaging with high fidelity. In addition, using 2,923 nm light, we demonstrate resolved-sideband cooling of tweezer-trapped strontium. Beyond enabling studies of collective emission phenomena in flexible arrangements of atoms, our platform opens novel opportunities for dipolar many-body physics and enhanced control over Rydberg dynamics and the strontium fine-structure qubit.
\end{abstract}

\maketitle

\section{Introduction}
Strontium (Sr) has emerged as a cornerstone for quantum science with neutral atoms~\cite{Ye2008quantum}, enabling accurate atomic clocks~\cite{Takamoto2005optical, Bloom2014optical, bothwell2022resolving}, continuous coherent matter waves~\cite{Chen2022continuous}, cavity-mediated collective photon emission~\cite{Norcia2018cavity}, probes of fundamental physics~\cite{Kondov2019molecular}, and platforms for many-body quantum physics~\cite{Buob2024strontium}. In recent years, the combination of optical tweezers and strontium~\cite{Norcia2018microscopic, Cooper2018alkaline} has led to groundbreaking experiments in quantum metrology~\cite{Madjarov2019, Young2020half, Shaw2024multi}, quantum simulation~\cite{Young2022tweezer, Shaw2024benchmarking}, and quantum computation~\cite{Madjarov2020high, Scholl2023erasure}. This broad range of capabilities is rooted in strontium's singlet–triplet level structure, which comprises broad, narrow, and ultra-narrow transitions.

\begin{figure}[h!]
\centering
\includegraphics[width=\columnwidth]{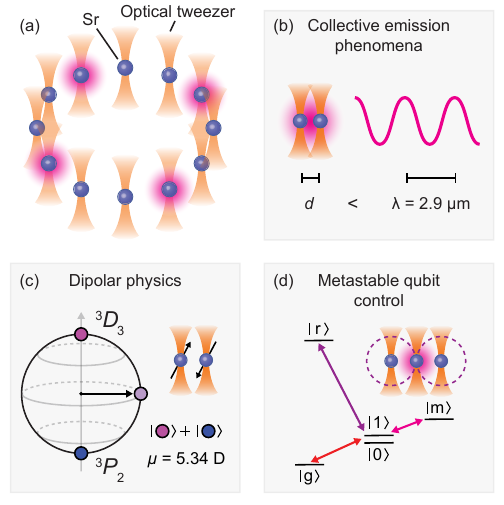}
\caption{Experimental potential of the strontium $2.9$ \textmu m transition in atomic tweezer arrays. (a) Optical tweezers, at a magic wavelength for the $2.9$ \textmu m transition, allow arrangements of single atoms with programmable geometries. (b) Subwavelength arrays, where the distance $d$ is smaller than the emission wavelength $\lambda$, open a path to studying collective emission phenomena. (c) The transition dipole moment of 5.34 Debye can enable new systems with elastic dipole-dipole interactions, realizing effective spin-spin coupling. (d) Coupling of $2.9$ \textmu m light to the fine-structure qubit opens novel opportunities for cooling and quantum state control.}
\label{fig:1}
\end{figure}

Central to strontium’s unique properties is the $5s5p\:^{3}P_{J}$ manifold of metastable states: ${}^{3}P_0$ is used as an ultranarrow clock state~\cite{Takamoto2005optical, Bloom2014optical}, ${}^{3}P_1$ is used for narrow-line optical cooling~\cite{Katori1999magneto}, and, more recently, ${}^{3}P_2$ has been explored as part of a fine-structure qubit~\cite{Unnikrishnan2024coherent, Pucher2024fine, Ammenwerth2025realization}. Transitions between the metastable $5s5p\,{}^{3}P_J$ states and the $5s4d\,{}^{3}D_{J^{\prime}}$ manifold are in the mid-infrared (mid-IR) range. The ${}^{3}P_0 \rightarrow{}^{3}D_1$ transition, at a wavelength of 2,603 nm, has been studied for its role in limiting the accuracy of strontium optical clocks via blackbody-induced losses~\cite{Nicholson2015systematic, Aeppli2024clock}. The $^{3}P_{2} \rightarrow\, ^{3}D_{3}$ transition at $\lambda = 2,923$~nm~\cite{Sansonetti2010wavelengths, Hashiguchi2019frequency} stands out with a set of particularly intriguing properties: its long wavelength comes with an unusually low recoil temperature of 12 nK, its linewidth is narrow at $57$~kHz~\cite{Safronova2013blackbody}, corresponding to an excited state lifetime of 2.8~\textmu s, and it allows optical cycling. In recent demonstrations with bulk Sr gases, this transition has been used to realize mid-IR magneto-optical traps~\cite{Hobson2020midinfrared, Akatsuka2021three} and sub-Doppler cooling schemes~\cite{Chen2025narrow}.

In combination with optical tweezer arrays, the 2.9~\textmu m transition promises to open a broader range of applications. The transition enables subwavelength atomic arrays, providing a path to studying collective emission phenomena, such as super- and subradiance. Importantly, it will be possible to realize subwavelength atomic arrays with flexible geometries~\cite{AsenjoGarcia2017exponential, Masson2022universality}, constituting a critical extension of recent experiments on collective phenomena using optical lattices with fixed square geometries~\cite{Rui2020Subradiant, Hutson2024Observation, Lu2025Supression, Douglas2026ManyBody}. In addition, the unusually large transition dipole moment can enable novel many-body systems with elastic dipole-dipole interactions~\cite{Olmos2013long} and the long wavelength can support enhanced cooperativity in atom-cavity interfaces~\cite{Covey2019telecom}. However, the 2.9~\textmu m transition has not previously been controlled with the spectral resolution, precision, and single-atom sensitivity required for tweezer-based experiments.

\begin{figure}[]
\centering
\includegraphics[width=\columnwidth]{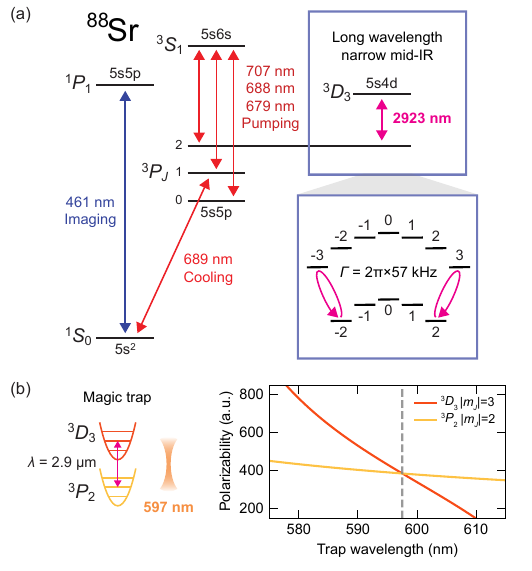}
\caption{Strontium level structure and trapping wavelength. (a) Level diagram of $^{88}$Sr and relevant transitions. The mid-infrared 2,923 nm transition from $^{3}P_{2}$ to $^{3}D_{3}$ is central to this work. The $m_{J}=\pm2\rightarrow \pm3$ transitions form closed two-level systems. (b) (left) Under magic trapping conditions, the optical tweezer potentials for atoms in the $^{3}P_{2}$ and $^{3}D_{3}$ states are identical. (right) Theoretical prediction of the polarizability for the $m_{J}=\pm2\rightarrow \pm3$ transitions, indicating magic conditions at around 597.5 nm.}
\label{fig:2}
\end{figure}

Here we demonstrate control of the $5s5p\:^{3}P_{2} \rightarrow\, 5s4d\:^{3}D_{3}$ 2.9 \textmu m transition with tweezer-trapped $^{88}$Sr atoms. We demonstrate magic trapping at 597.14(3) nm, preparation of single atoms, and site-resolved imaging across the array with high fidelity. In this setting, we perform precision spectroscopy of the 2.9 \textmu m line, resolving sublevels at the level of the natural linewidth using a kHz-resolution mid-IR laser system. Enabled by the narrow linewidth, we also achieve resolved-sideband cooling, bringing tweezer-trapped atoms in the $^{3}P_{2}$ metastable state close to their motional ground state.

\begin{figure}[]
\centering
\includegraphics[width=\columnwidth]{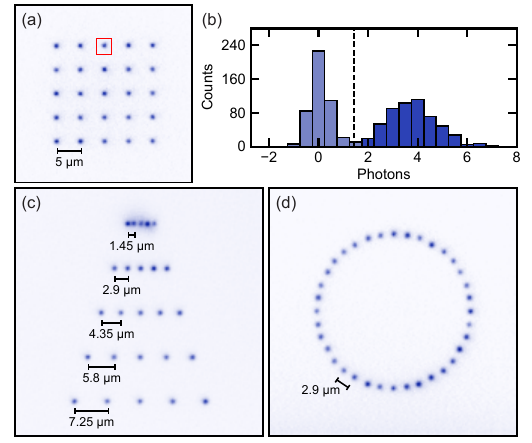}
\caption{Single-atom preparation and imaging in 597 nm tweezers. (a) An averaged image of a $5\times5$ atomic tweezer array with 5 \textmu m spacing. (b) Back-to-back imaging analysis for a typical trap, highlighted with a red box in panel (a). (c) Averaged fluorescence image of one-dimensional chains with $\lambda/2$, $\lambda$, $3/2\lambda$, $2\lambda$ and $5/2\lambda$ spacing. (d) Averaged fluorescence image of a ring-shaped array with 36 sites spaced by $\lambda$. Each image in (a), (c), and (d) is an average of 100 individual images without parity projection.}
\label{fig:3}
\end{figure}

\begin{figure*}[]
\centering
\includegraphics[width=\textwidth]{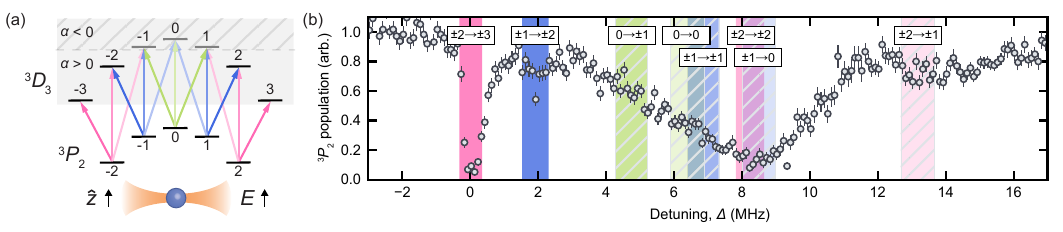}
\caption{Spectroscopy of the 2.9 \textmu m $^{3}P_{2}\rightarrow{}^{3}D_{3}$ transition in a Sr tweezer array at 597 nm. (a) An energy level diagram illustrating the shifts induced by 597 nm light on the Zeeman sublevels of the $^{3}P_{2}$ and $^{3}D_{3}$ manifolds; levels in the solid gray (hatched gray) are trapped (anti-trapped). The tweezer light is linearly polarized. (b) Loss spectrum of tweezer-trapped atoms following illumination with 2.9 \textmu m light. Here, $\Delta$ is defined as the detuning relative to the free-space resonance. Vertical color bars indicate the predicted resonance shifts using the same color scheme as in (a); hatched color bars indicate transitions to anti-trapped levels.}
\label{fig:4}
\end{figure*}

\section{Platform for mid-IR Experiments}
A Sr mid-infrared platform building on the 2.9 \textmu m transition offers intriguing experimental possibilities, as illustrated in Fig.~\ref{fig:1}. Subwavelength arrays enable studies of collective photon emission~\cite{AsenjoGarcia2017exponential, Bettles2016enhanced, Shahmoon2017cooperative, Masson2022universality} and quantum metasurfaces~\cite{Perczel2017topological, Bekenstein2020quantum}, with applications to single-photon storage and retrieval~\cite{AsenjoGarcia2017exponential, Manzoni2018optimization, Ballantine2021quantum}, dark-state engineering~\cite{RubiesBigorda2022photon, Ballantine2021quantum}, and collectively enhanced cooling~\cite{RubiesBigorda2025collectively}. The transition dipole moment of $5.34$ Debye, comparable to strongly dipolar molecules~\cite{Bigagli2024observation}, opens routes to novel many-body physics with elastic dipole-dipole interactions, such as quantum spin systems and models of quantum magnetism. In conjunction with cooperative emission, such systems may realize complex open quantum systems in a controlled way~\cite{Brennen1999quantum,Olmos2013long}. From the metastable $^3P_2$ state, atoms can also be coupled to highly excited Rydberg states, complementing fine-structure qubits~\cite{Unnikrishnan2024coherent, Pucher2024fine, Ammenwerth2025realization} and enabling control over inelastic dipolar interactions~\cite{Srakaew2023subwavelength}. Laser cooling on this transition could also reduce Doppler-induced dephasing in Rydberg gates, enable studies of spin-motion entanglement~\cite{Graham2019rydberg, Zhang2024motional}, and improve the performance of tweezer clocks~\cite{Chen2025narrow}.
Finally, the resonant scattering cross section of an electric dipole transition scales $\propto \lambda^2$, which can be advantageous for atom--photon coupling in cavity and waveguide QED systems.

In this work we use $^{88}\text{Sr}$; the relevant levels and transitions are shown in Fig.~\ref{fig:2}(a). The broad line at 461 nm, the narrow intercombination line at 689 nm, and repumpers at 679 nm, 688 nm, and 707 nm are used for cooling, state preparation, and imaging. The mid-IR transition at 2,923 nm between $5s5p\:^{3}P_{2}$ and $5s4d\:^{3}D_{3}$ is at the center of this study. As illustrated in Fig.~\ref{fig:2}(a), it is comprised of several sublevels. Importantly, transitions between the stretched states, $m_J = + 2 \rightarrow + 3$ and $- 2 \rightarrow - 3$, form closed two-level systems due to selection rules.

Our experimental setup allows the interrogation of ultracold, tweezer-trapped strontium atoms with narrow 2.9 \textmu m laser light. Details on the cooling of Sr atoms in our setup have been reported in earlier work~\cite{Kwon2023jet, Holman2026trapping}. The optical tweezer array, which is generated via holographic metasurfaces~\cite{Huang2023metasurface, Holman2026trapping} (see Appendix for additional details), is loaded from a gas of Sr atoms in the $^1S_0$ state at microkelvin temperatures. Single-mode and single-frequency laser light at a wavelength of 2.9 \textmu m is produced via difference frequency generation (DFG) using 1578 nm and 1025 nm seed lasers. By stabilizing the seed lasers to an ultralow-expansion cavity, we reach a linewidth of less than 10 kHz for the 2.9 \textmu m light. The laser system provides a total optical power of approximately 150 mW of which we only use a small fraction for the experiments in this work. Due to the long wavelength and narrow linewidth, the saturation intensity ($I_{\mathrm{S}}$) of the 2.9 \textmu m transition is as small as $0.30 \text{ \textmu W/cm}^2$.

A typical experimental sequence includes the following steps: The tweezer-trapped atoms are optically pumped from the $^1S_0$ state to the metastable $^{3}P_{2}$ state using the 689 nm, 688 nm and 679 nm transitions (see Appendix for additional details). Due to its long lifetime of about 500 seconds~\cite{Yasuda2004lifetime}, $^{3}P_{2}$ can be viewed as a de facto ground state compared to the time scale of typical experiments (about 1 second). Then, the atoms in $^{3}P_{2}$ are exposed to 2.9 \textmu m laser light and coupled to the $^{3}D_{3}$ state. To read out the remaining population in $^{3}P_{2}$, we transfer atoms back to the absolute ground state $^1S_0$ via optical pumping on the 707 nm and 679 nm transitions and spontaneous decay from $^{3}P_{1}$ in less than 1 ms. Finally, we perform fluorescence imaging via 461 nm light.

\section{Single-atom preparation and imaging}
We demonstrate trapping of single Sr atoms and high-fidelity detection at the previously unexplored tweezer wavelength at 597 nm. The choice of this wavelength is motivated by the prediction of magic trapping conditions for the closed $m_J = \pm 2 \rightarrow \pm 3$~\cite{Kiruga2025portal} transitions [see Fig.~\ref{fig:2}(b)]. Besides being magic, this wavelength traps the relevant intermediate electronic states of Sr and has a relatively short wavelength in the visible, enabling diffraction-limited tweezer spots at close spacings and arrays in the subwavelength regime.

The optical tweezers are linearly polarized and the polarization vector is aligned with the vertical $z$-direction, along the direction of gravity. Atoms are loaded into the tweezers by ramping up the trap depth to $k_\mathrm{B} \times 0.08$ mK for 50 ms. Here, $k_\mathrm{B}$ denotes the Boltzmann constant. Each trap is initially loaded with $\langle N \rangle \approx 2$ atoms on average. Fig.~\ref{fig:3}(a) shows an averaged image of a $5\times5$ tweezer array. Single atoms are then prepared by parity projection, using light-assisted collisions with 689 nm light~\cite{Zelevinsky2006narrow} to induce pairwise loss and project even occupation to $N=0$ and odd occupation to $N=1$ (see Appendix).

To determine the atomic occupation of each trap, we perform fluorescence imaging on the 461 nm transition with simultaneous repulsive Sisyphus cooling on the 689 nm transition. Scattered photons are collected with a high-NA objective and recorded on a single-photon-sensitive camera. Atoms are imaged for 60 ms in a $k_\mathrm{B} \times 3$ mK trap; each trap is classified as empty or occupied based on the detected photon number. The histogram of collected photons, see Fig.~\ref{fig:3}(b), shows well-separated photon-count distributions for zero and one atom per trap. For a typical trap, we measure an imaging fidelity of 99.7(3)\%, a filling rate of 54.7(5)\%, and a survival rate of 90.3(4)\% for one imaging cycle (see Appendix for details on imaging characterization).

We have trapped atoms in various array geometries, using different holographic metasurfaces~\cite{Holman2026trapping}. The realized atomic arrays include one-dimensional lines with spacings in the range from 1.45 \textmu m to 7.25 \textmu m as shown in Fig.~\ref{fig:3}(c), as well as a ring with atoms spaced at distance $\lambda$, as shown in Fig.~\ref{fig:3}(d). Such arrangements will be useful for studies of cooperative emission effects in unity-filled arrays. All data in the remainder of this work is recorded with atoms trapped in a $5 \times 5$ array [see Fig.~\ref{fig:3}(a)]. At a trap spacing of 5 \textmu m, the atoms are placed sufficiently far apart to suppress spurious collective effects and minimize elastic dipole-dipole interactions.

\begin{figure}[]
\centering
\includegraphics[width=\columnwidth]{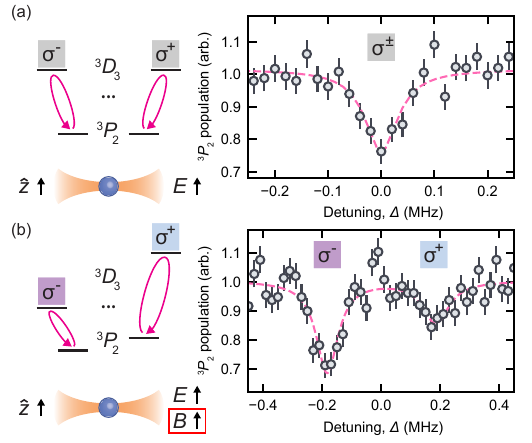}
\caption{Precision spectroscopy of the closed $|m_{J}|=2\rightarrow 3$ transition. (a) In the absence of a magnetic field, a linewidth of $\Gamma/(2 \pi) = 74(10)$ kHz is measured, obtained from a Lorentzian fit. (b) In the presence of a magnetic field the previously degenerate transitions $m_{J}= \pm 2\rightarrow \pm 3$ split symmetrically. The splitting is about 0.4 MHz. The magnetic field is about 0.14(1) G and aligned along the polarization vector of the optical tweezers. The linewidth of the left dip and the right dip are $\Gamma/(2 \pi) = 90(15)$ and $107(20)$ kHz, respectively.}
\label{fig:5}
\end{figure}

\section{State-resolved spectroscopy}
Using tweezer-trapped single atoms, we perform spectroscopy on the 2.9 \textmu m transition. First, we apply 2.9 \textmu m light for 50 ms, with $I \gg I_\mathrm{S}$. It is important to note that for the trapping wavelength of 597 nm, the sublevels of $^{3}D_{3}$ differ in their optical polarizability. As highlighted in Fig.~\ref{fig:4}(a), some sublevels have a polarizability $\alpha > 0$ ($\alpha < 0$), corresponding to a trapping (anti-trapping) tweezer potential. As a result, there are two distinct loss mechanisms following illumination with 2.9 \textmu m light: sublevels that are trapped ($|m_{J}|=2,3$) experience recoil heating, while sublevels that are anti-trapped ($|m_{J}|=0,1$) experience immediate atom expulsion when interacting resonantly with 2.9 \textmu m light. In the absence of external fields, all transitions between $^{3}P_{2}$ and $^{3}D_{3}$ sublevels are degenerate, while in the presence of trap light, the atoms experience a light shift that lifts the degeneracy. The resulting level shifts are illustrated in Fig.~\ref{fig:4}(a).

By scanning the frequency of the 2.9 \textmu m laser, we record an in-trap loss spectrum of the $^{3}P_{2}\rightarrow{}^{3}D_{3}$ transitions, as shown in Fig.~\ref{fig:4}(b). The peak intensity of each tweezer is $I_0 = 0.34(6)~\mathrm{MW/cm^2}$. This intensity is calculated from the total optical power delivered to the array, the number of tweezers, and a tweezer waist of $1.1 \pm 0.1$ \textmu m; the latter is inferred from parametric trap-frequency measurements. The spectral structure of the observed loss features aligns with the predicted locations from the theoretical polarizabilities~\cite{Kiruga2025portal}. Around the predicted locations for transitions to anti-trapped excited states we find broad loss features. We attribute these broad features to off-resonant scattering into the continuum of unbound motional states. At the low frequency end of the spectrum, separated from the other transitions, we observe a deep and narrow feature close to the location expected for the $|m_{J}|=2 \rightarrow 3$ transition.

Next, we focus on the $|m_{J}|= 2 \rightarrow 3$ transition and perform precision spectroscopy with $I \lesssim I_\mathrm{S}$. We measure a linewidth of $\Gamma/(2 \pi) = 74 \pm 10$ kHz, as shown in Fig.~\ref{fig:5}(a). Here both $\pm \sigma$ transitions are driven simultaneously. Compared to the expected linewidth of 57 kHz, we attribute the broadening to intensity fluctuations of the probe laser on the order of the very small saturation intensity $I_\mathrm{S}$. To lift the degeneracy between the $m_{J}= \pm 2\rightarrow \pm 3$ transitions, we apply a magnetic field oriented along the polarization axis of the tweezers. This induces a Zeeman splitting between the two transitions as shown in Fig.~\ref{fig:5}(b). The different depths of the observed dips originate from an imbalance between the initial atomic population in $m_{J}= \pm 2$ sublevels of $^{3}P_{2}$, which depends on details of the state preparation via optical pumping. Compared to the data in Fig.~\ref{fig:5}(a), a slight broadening of the linewidth is attributed to an increased probe laser intensity.

\begin{figure}[]
\centering
\includegraphics[width=\columnwidth]{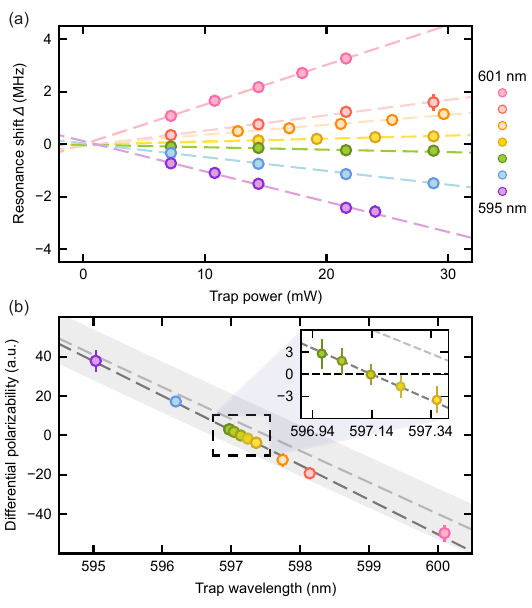}
\caption{Determining the magic wavelength for the $|m_{J}|=2\rightarrow 3$ transition. (a) Resonance shift as a function of optical power per tweezer trap. Dotted lines are linear fits with the intercept point corresponding to the free-space resonance. (b) Differential polarizability as a function of wavelength. The data points are obtained from fitting the slopes in panel (a) and the measured trap waist. The dashed light gray line is the theoretical prediction and the shaded region around it indicates its uncertainty. The dashed dark gray line is a linear fit to the experimental data. The fit yields a slope of $-17.5 \pm 1$ a.u./nm and a zero crossing of the differential polarizability at $597.14 \pm 0.03$ nm. The inset shows a fine scan of wavelengths close to the zero crossing.}
\label{fig:6}
\end{figure}

\section{Magic wavelength measurement}
Precision spectroscopy on the 2.9 \textmu m transition allows us to accurately determine the tweezer wavelength for which magic trapping conditions are achieved. Specifically, we are interested in magic conditions for linearly polarized trap light, which ensures that the $|m_J|=2 \rightarrow 3$ transition is a closed 2-level system. Magic trapping, where the differential light shift between two states vanishes, is critically important for future applications that rely on coherent control of the 2.9 \textmu m transition.

We measure the differential light shift of the 2.9 \textmu m transition as a function of the trapping wavelength. For several trapping wavelengths in the range 595--600 nm, we vary the trap intensity, measure the corresponding in-trap resonance frequency, and observe the resonance shift which generally depends on trap intensity [see Fig.~\ref{fig:6}(a)]. For each wavelength, we apply a linear fit to the data and extract the differential polarizability from the slope of the fit. Fig.~\ref{fig:6}(b) shows the measured differential polarizability as a function of trap wavelength, together with the theoretical prediction, obtained using an approach that combines configuration interaction with the coupled cluster method~\cite{Cheung2025pCI}. The experimental data is well within the uncertainty of the prediction, slightly offset from the theoretical mean value. This measurement will help to further refine the precise spectroscopic model of strontium~\cite{Kiruga2025portal}. A fine scan, shown in the inset of Fig.~\ref{fig:6}(b), narrows down the location of the magic wavelength to $597.14 \pm 0.03$ nm.

\section{Sideband cooling}
\begin{figure}
\centering
\includegraphics[width=\columnwidth]{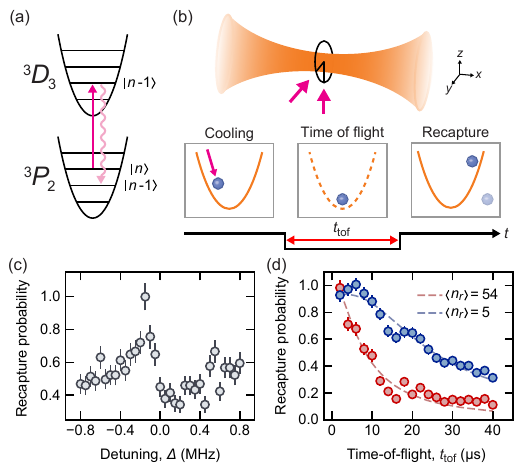}
\caption{Resolved-sideband cooling on the 2.9 \textmu m transition. (a) Schematic of resolved-sideband cooling in magic wavelength tweezers. Driving the spectrally resolved red sideband, on average, lowers the atom's motional state. (b) Illustration of a tweezer trap and the directions along which 2.9 \textmu m cooling light is applied. Following cooling, the tweezers are turned off for a variable amount of time, $t_\mathrm{tof}$, during which the atoms expand according to a velocity distribution that is set by their temperature. (c) Recapture probability as a function of 2.9 \textmu m detuning. Maximum survival probability is observed for a laser frequency that is red-detuned by approximately 150 kHz. The radial trap frequencies are $\nu_\mathrm{r}=95(5)$ kHz. (d) Time-of-flight measurements for atoms initialized in $^{3}P_{2}$  without cooling (red data) and after cooling (blue data). Dashed lines correspond to fits of a classical Monte Carlo simulation assuming atoms in free flight sampled from a Boltzmann distribution.}
\label{fig:7}
\end{figure}

We demonstrate direct sideband cooling of strontium in the metastable $^3P_2$ state using the 2.9 \textmu m mid-IR transition. This becomes possible as three key conditions are fulfilled: (1) Due to the narrow linewidth of the 2.9 \textmu m transition of less than 100 kHz and typical radial trap frequencies of about 100 kHz, motional sidebands are resolved. (2) Owing to the tight confinement and the long transition wavelength, the system is deep in the Lamb-Dicke regime. The Lamb-Dicke parameter in our system is $\eta \approx 0.07$. (3) The atoms are trapped under magic conditions, such that resonance frequencies between motional states are identical in each trap. Fig.~\ref{fig:7}(a) illustrates the resolved-sideband cooling mechanism for atoms in the $^3P_2$ state, using the 2.9 \textmu m transition to $^3D_3$.

In order to characterize the cooling efficiency, we use a release-and-recapture sequence that measures the effective kinetic energy of tweezer-trapped atoms. After cooling, the atoms are released from the tweezer for a duration $t_\mathrm{tof}$ before the tweezers are switched on again for recapture. The sequence is illustrated in Fig.~\ref{fig:7}(b). Cooling is performed with two 2.9 \textmu m beams along the radial $y$ and $z$ directions. The cooling light is pulsed on and off one hundred times with a cycle time of 130 \textmu s (90 \textmu s on, 40 \textmu s off). The dark time allows population, which is excited during each pulse, to return to $^{3}P_{2}$ before the next cooling pulse. The pulse durations are chosen empirically to minimize the atomic kinetic energy.

We first keep $t_\mathrm{tof}$ fixed at 6 \textmu s, vary the detuning of the 2.9 \textmu m laser, and observe an optimal detuning for which the recapture probability is maximal, as shown in Fig.~\ref{fig:7}(c). A higher recapture probability indicates lower temperatures of the trapped atoms in the $5 \times 5$ array. We observe a maximum in the recapture probability for a laser frequency that is red-detuned by about 150 kHz from the free-space resonance. At this maximum the recapture probability is significantly enhanced compared to the case where 2.9 \textmu m cooling light is absent, indicating motional sideband cooling. The optimal detuning is slightly larger than the measured trap frequency, which we attribute to power broadening: a more red-detuned beam reduces off-resonant carrier scattering that would otherwise cause heating.

In order to extract the temperature, we measure the recapture probability as a function of time-of-flight and compare the results to a classical Monte Carlo simulation, as shown in Fig.~\ref{fig:7}(d). We first measure the initial temperature of atoms prepared in the $^{3}P_{2}$ state, before the cooling sequence is applied. For the classical Monte Carlo model the three-dimensional velocities of trapped atoms are drawn from a finite temperature Boltzmann distribution that is homogeneous across all three dimensions (see Appendix). Without cooling, a fit with the Monte Carlo model yields an average temperature of 126(10) \textmu K, corresponding to an average radial motional state of $\expval{n_r}=54 \pm 5$.

In the presence of radial cooling with 2.9 \textmu m light we observe a substantial increase in recapture probability (see Fig.~\ref{fig:7}(d), blue data). To extract the radial temperature, we fit the experimental data with the same classical Monte Carlo model with the modification that the axial temperature is kept constant, reflecting the absence of cooling in the axial $x$-direction (see Appendix for further details). After cooling, we measure an average radial temperature of 14(5) \textmu K, corresponding to an average radial motional state of $\expval{n_r}=5 \pm 2$. This corresponds to a reduction in temperature compared to the uncooled system by about an order of magnitude, bringing atoms close to their radial ground state. We attribute the residual radial excitation in part to heating along the axial direction, as we cannot completely isolate the weakly confined axial mode. As the axial temperature is assumed to be constant in the model, the model effectively yields a higher radial temperature, and as such constitutes a conservative upper bound. Efficient sideband cooling on the 2.9 \textmu m transition establishes a new route for motional control of tweezer-trapped atoms in the ${}^3P_2$ state.

\section{Outlook}

We have realized a strontium tweezer platform that enables the use of the mid-IR 2.9 \textmu m transition for a broad range of quantum applications. We have achieved high-fidelity trapping and imaging of single atoms in  optical tweezer arrays that are magic for the 2.9 \textmu m transition, and demonstrated sideband cooling of tweezer-trapped Sr atoms in the metastable $^3P_2$ state using 2.9 \textmu m light.

In a next step, our platform will enable coherent quantum state control on the $^{3}P_{2} \rightarrow\, ^{3}D_{3}$ transition via narrow 2.9 \textmu m light. Combined with straightforward improvements to achieve high-fidelity internal state preparation via controlled optical pumping and the preparation of defect-free arrays, our platform opens a path towards subwavelength optical tweezer arrays that promise the investigation of cooperative emission effects in novel regimes. Experiments that come within reach include the observation of collective level shifts in one-dimensional subwavelength chains~\cite{Hofer2025single}, the observation of Dicke superradiance~\cite{Masson2024dicke}, the engineering and control of subradiant states~\cite{AsenjoGarcia2017exponential, RubiesBigorda2022photon}, and other exotic radiative effects of correlated quantum emitters~\cite{Alaee2020Kerker, Agarwal2024entanglement, Holzinger2024harnessing}.

Our work also enhances broader experimental capabilities for strontium tweezer arrays. From the $^3P_2$ state, atoms can be coupled to highly excited Rydberg states, enabling studies that combine Rydberg blockade effects and the fine-structure qubit~\cite{Unnikrishnan2024coherent, Pucher2024fine, Ammenwerth2025realization}. The sideband cooling demonstrated here may improve Rydberg-mediated gates by suppressing Doppler-induced dephasing and enabling controlled studies of spin-motion coupling~\cite{Graham2019rydberg, Zhang2024motional}. Finally, the techniques demonstrated here may also support research on Sr optical tweezer clocks~\cite{Madjarov2019, Young2020half}.

\section*{Acknowledgments}
We thank Henri Dhonte and Benjamin Czasch for experimental assistance and Ana Asenjo-Garcia, Stuart Masson, and Cosimo Rusconi for fruitful discussions. We also acknowledge helpful discussions with Ana Maria Rey and Susanne Yelin. This work was supported by the National Science Foundation (Award nos.~1936359, 2040702, and 2004685) and the Air Force Office of Scientific Research (Award nos.~FA9550-16-1-0322 and FA9550-23-1-0404). Metasurface fabrication was carried out at the Columbia Nano Initiative cleanroom and at the Advanced Science Research Center Nanofabrication Facility at the Graduate Center of the City University of New York. N.Y.~acknowledges the Gordon and Betty Moore Foundation Experimental Physics Investigators (EPI) initiative (Award no.~11561). Theory work was supported by Office of Naval Research (Award no.~N000142512105), National Science Foundation QLCI (Award no.~2016244), and National Science Foundation (Award no.~2309254). S.W. acknowledges support from the Alfred P. Sloan Foundation.

{\bf Author contributions.}
All authors contributed substantially to the work presented in this paper. A.H., X.S., B.S., and J.C.~carried out the atomic experiments. Z.Z., Y.X., and J.W.~designed and fabricated the metasurfaces. N.Y.~supervised the metasurface work. D.F.~and M.S.~provided theoretical atomic data. S.W.~supervised the study. A.H., X.S., and S.W.~analyzed and interpreted the data and wrote the manuscript, with input from all authors. \\

{\bf Data availability.}
The experimental data that supports the findings of this study are available from the corresponding author upon reasonable request. \\

{\bf Competing interests.}
The authors declare no competing interests. \\

\clearpage


\appendix

\section{\MakeUppercase{Experimental setup}}
\begin{figure}[h]
\centering
\includegraphics[width=\columnwidth]{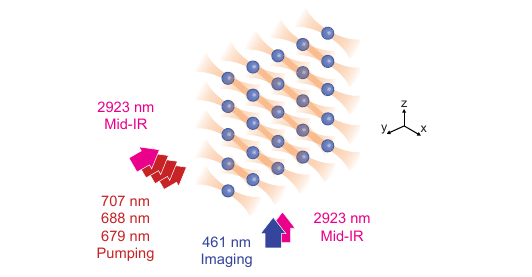}
\caption{Laser beams for cooling and state preparation of $^{88}$Sr relative to the orientation of the optical tweezer array.}
\label{fig:A0}
\end{figure}

The $^{88}$Sr atoms in the optical tweezer array are controlled by several laser beams as illustrated in Fig.~\ref{fig:A0}. For cooling, state preparation, and imaging, the broad 461 nm ($\Gamma = 2\pi\times 30$ MHz), the narrow intercombination line at 689 nm ($\Gamma = 2\pi\times 7.4$ kHz), and pumping light at 679 nm, 688 nm, and 707 nm ($\Gamma = 2\pi\times 1.3$ MHz, $\Gamma = 2\pi\times 3.9$ MHz, and $\Gamma = 2\pi\times 7.2$ MHz) are used. The 2,923 nm ($\Gamma = 2\pi\times 57$ kHz) light is sent in along the $y$ for spectroscopy, and along the $y$ and $z$ for sideband cooling to address the radial sidebands. The polarization of the $y$ ($z$) beam is linear along the $x$ ($y$). The beams drive both $\sigma \pm$ transitions of the atoms.

\section{\MakeUppercase{Single-atom preparation and imaging}}

Initially a mean number of $\langle N \rangle \approx 2$ atoms is loaded into each tweezer trap. Single atoms are prepared via parity projection by applying 689 nm light red-detuned by 300 kHz from the in-trap $^{1}S_{0} \rightarrow\, ^{3}P_{1},\, m_{J}=0$ resonance for 50 ms. For imaging, atoms are trapped at a trap depth of 3 mK and resonant 461 nm light is applied for 60 ms. Each trap is evaluated to have no atom (few photons detected) or one atom (many photons detected). Fluorescence photons are collected via a high NA objective (NA = 0.55) and recorded with a low-noise, single-photon sensitive camera. During imaging, a repulsive Sisyphus cooling scheme on the $^{1}S_{0} \rightarrow\, ^{3}P_{1},\, m_{J}=0$ transition counteracts heating from scattered imaging photons.

\begin{figure}[]
\centering
\includegraphics[width=\columnwidth]{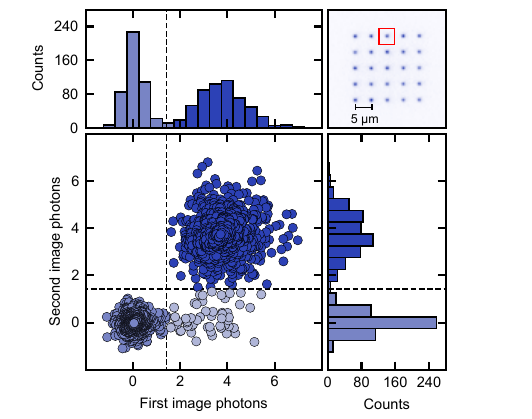}
\caption{Single-atom imaging performance. Back-to-back photon counts for a typical trap in the array (highlighted by the red box). The threshold value $x$ divides the data into four quadrants, indicating the absence or presence of an atom in the first and second image. The top-left and bottom-right panels show the histograms of each image's photon counts.}
\label{fig:A1}
\end{figure}

To characterize the imaging performance and losses during imaging, we take two consecutive images and compare the detected photon numbers in 1,000 iterations of the experiment. A model-free approach that makes no assumptions about the functional shape of the distribution of the collected photons allows us to determine the imaging fidelity~\cite{Norcia2018microscopic, Cooper2018alkaline}. A detailed description of the approach can be found in Ref.~\cite{Holman2026trapping}. Fig.~\ref{fig:A1} shows 2D histograms of the recorded photons from the consecutive images for the trap that is highlighted by a red box. The corresponding single atom detection fidelity is 99.7(3)\%.

\section{\MakeUppercase{Metasurface-generated traps}}

The optical tweezer arrays in Fig.~\ref{fig:3} are produced by a transmissive, phase-only, holographic metasurface composed of TiO$_2$ nanopillars patterned on a fused-silica substrate. Both the holographic phase profile and the library of TiO$_2$ nanopillars are designed for operation at the 597-nm tweezer wavelength. The phase profile is optimized for optical efficiency and for intensity uniformity of the tweezer array using a weighted Gerchberg-Saxton algorithm. The phase response and the forward scattering efficiency (i.e., amplitude response) of the nanopillars are calculated using rigorous coupled-wave analysis with periodic boundary conditions. The optimized phase profile is implemented by a 2D square lattice of nanopillars with desired phase response and sufficiently high amplitude response. The nanopillars have a uniform height of 600 nm and square cross-sections with variable sizes ranging from 90 nm to 260 nm; the pitch of the nanopillar lattice is 360 nm.

The metasurfaces are fabricated using a process described in Ref.~\cite{Holman2026trapping}. A resist template defined by electron-beam lithography on a fused-silica substrate is conformally coated with TiO$_2$ using atomic layer deposition; inductively coupled plasma reactive-ion etching is then used to remove excess TiO$_2$ down to the resist surface; O$_2$ plasma ashing is finally used to remove the resist template, leaving free-standing TiO$_2$ nanopillars on the fused silica substrate. In the experiment, the metasurface converts a normally incident collimated beam at 597 nm directly into a tweezer array, which is projected into the glass cell with a one-to-one relay optic system to trap Sr atoms.

\section{\MakeUppercase{$^3P_2$ state preparation}}

To optically pump atoms from the ground state to the $^3P_2$ state, the 689 nm, 688 nm, and 679 nm lasers are simultaneously turned on for 10 ms. By taking advantage of selection rules between the intermediate states, in particular the decay channels from $^3S_1$, atoms can be predominantly prepared in $|m_{J}| = 2$ state of the $^3P_2$ manifold.

Specifically, 689 nm light is tuned on resonance with the in-trap $\sigma\pm$-transition. Owing to the different polarizabilities of $|m_{J}| = 1$ and $m_{J} = 0$ in $^3P_1$, a majority of the ground state population is brought to $^3P_1, |m_{J}| = 1$. Atoms in $^3P_1$ are then pumped to $^3S_1, |m_{J}| = 1$ by the $\pi$-polarized 688 nm pump. When atoms decay from $^3S_1, |m_{J}| = 1$, the branching ratio greatly favors decay to $^3P_2$ (about $60\%$ of all $^3P_J$ states), especially $|m_{J}| = 2$ (about $60\%$ of all $|m_{J}|$ states). Atoms that decay to the $^3P_1$ state ($30\%$) reenter the pumping cycle, while atoms that decay to the $^3P_0$ ($10\%$) are driven by $\sigma\pm$ 679 nm pump back to $^3S_1, |m_{J}| = 1$. The pumping duration of 10 ms has been empirically optimized.

\begin{figure}
\centering
\includegraphics[width=\columnwidth]{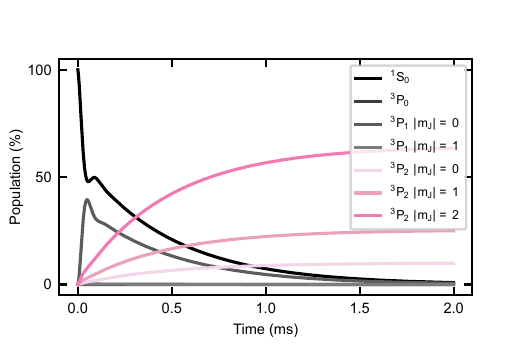}
\caption{Numerical results from a rate equation model for optical pumping from $^1S_0$ to $^3P_2$. Simultaneous application of 689~nm, 688~nm, and 679~nm pumping light transfers atoms from $^{1}S_{0}$ to the $^{3}P_{2}$ manifold, with population predominantly accumulating in the stretched \(|m_J|=2\) states.}
\label{fig:A2}
\end{figure}

To model the optical pumping dynamics, we use a numerical method to solve the multi-level master equation of our optical pumping process, accounting for optical powers and potentially imperfect polarization. The expected population distribution in the steady state can be inferred from the theoretical data in Fig.~\ref{fig:A2}.

\section{\MakeUppercase{Release and Recapture model}}
To extract the radial temperature in a tweezer trap, we fit the measured release-and-recapture curve to a classical Monte Carlo simulation of atoms released from the tweezers, closely following the model detailed in Ref.~\cite{Biagioni2025narrowline}. The tweezer is modeled as a Gaussian optical dipole trap, with a profile determined by the experimentally measured trap waist, optical power, and atomic polarizability.

We first use an isotropic temperature distribution across all axes to characterize the initial temperature of atoms in the tweezers. We observe atoms in the $^1S_0$ state to be at approximately 17(5)~\textmu K, and atoms optically pumped into the $^3P_2$ state to be at approximately 126(10)~\textmu K. Since we only cool atoms in the $^3P_2$ state on the radial sidebands, we assume that the axial temperature remains fixed at 126(10)~\textmu K. The release-and-recapture data after cooling are therefore fit with a temperature representing only the radial direction.

\bibliography{references}
\end{document}